\documentclass[floatfix,preprint, onecolumn, prd, showpacs,nofootinbib]{revtex4-1}
\usepackage{natbib}
\usepackage{times}
\usepackage{amssymb,amsbsy,amsmath,amsfonts}
\usepackage{graphicx}% Include figure files
\usepackage{float}
\usepackage{morefloats}
\usepackage{rotating}
\usepackage{srcltx}
\usepackage{slashed}
\begin{document}

\title{Scattering lengths of Nambu-Goldstone bosons off $D$ mesons  and dynamically generated heavy-light mesons}

\author{ M. Altenbuchinger$^1$, L.-S. Geng,$^{2,1}$   and W. Weise$^{1,3}$}
\affiliation{
$^1$Physik Department, Technische Universit\" at M\"unchen, D-85747 Garching, Germany\\
$^2$School of Physics and Nuclear Energy Engineering and International Research Center for
Particles and Nuclei in the Cosmos, Beihang University,  Beijing 100191,  China\\
$^3$ECT*, Villa Tambosi, I-38123 Villazzano (Trento), Italy}

 \begin{abstract}
Recent lattice QCD simulations of the scattering lengths of Nambu-Goldstone bosons off the $D$ mesons
are studied using unitary chiral perturbation theory.  We show that the Lattice QCD data are better described
in the covariant formulation than in the heavy-meson formulation. The  $D^*_{s0}(2317)$ can be dynamically generated from
 the coupled-channels $DK$ interaction without \textit{a priori} assumption of its existence.
A new renormalization scheme is proposed which manifestly satisfies chiral power counting rules and has well-defined behavior in the infinite heavy-quark mass limit.  Using this scheme we predict the heavy-quark spin and flavor symmetry counterparts of the $D^*_{s0}(2317)$.
\end{abstract}

\pacs{12.39.Fe,  13.75.Lb, 14.40.Lb,  14.40.Nd}

\date{\today}

\maketitle

\section{Introduction}
Measurements of hadronic states with charm quarks such as the $D_{s0}^*(2317)$ have led to extensive
and still ongoing discussions about our deeper understanding of mesons and baryons~\cite{Aubert:2003fg,Krokovny:2003zq,Besson:2003cp}, traditionally thought to be composed of a pair of quark and antiquark or three quarks in the naive quark model.
With its mass ($M=2317.8\pm0.6$ MeV) about 100 MeV lower than the  lowest $c\bar{s}$ scalar state in the naive quark model, the
$D^*_{s0}(2317)$ cannot be a conventional $q\bar{q}$ state~\cite{Bardeen:2003kt,Nowak:2003ra,vanBeveren:2003kd,Dai:2003yg,Narison:2003td,Chen:2004dy,Szczepaniak:2003vy,Browder:2003fk,Cheng:2003kg,Barnes:2003dj,Kolomeitsev:2003ac,Guo:2006fu,Gamermann:2006nm,Zhu:2007wz}. One possible interpretation is
that of a compound dynamically generated by the strong $DK$ interaction in coupled-channels  dynamics~\cite{Kolomeitsev:2003ac,
Guo:2006fu,Gamermann:2006nm}. Such approaches have provided many useful insights into the nature of some most intriguing new resonances (see, e.g., Refs.~\cite{Garcia-Recio:2013uva,Ozpineci:2013zas} for some recent applications).

In order to clarify the nature of the $D_{s0}^*(2317)$, or of any other meson of similar kind, it is useful
to study such objects from various perspectives and compare the results with experimental and lattice QCD (LQCD) data. In this respect, it has been argued that  the isospin-breaking decay width $D_{s0}^*(2317)\rightarrow D_s\pi$~\cite{Lutz:2007sk,Guo:2008gp},
the light-quark mass dependence~\cite{Cleven:2010aw},  and the volume dependence~\cite{MartinezTorres:2011pr} of the $D_{s0}^*(2317)$ properties can
provide valuable information on its nature. At the same time it should also be noted that, in addition to
the $D_{s0}^*(2317)$, coupled-channels unitary dynamics predicts several other states in sectors or channels related to the $D_{s0}^*(2317)$ by heavy-quark spin and flavor symmetry
and (approximate) chiral symmetry [or broken SU(4) symmetry]~\cite{Kolomeitsev:2003ac,Guo:2006fu,Gamermann:2006nm,  Hofmann:2003je,Guo:2006rp}. Once the mass and width of the $D_{s0}^*(2317)$ are fixed, so are those of the other related states. Future experiments in search for those resonances in the predicted energy regions are therefore strongly encouraged.

All these predictions are subject to potentially sizable symmetry-breaking corrections. In particular,  a comprehensive study of recoil corrections  is necessary because the velocity of the charm quark in $D(D^*)$ mesons is
only about $0.3 c$, not small enough to allow for a complete  neglect of recoil corrections. For the scattering lengths of  the Nambu-Goldstone bosons off the $D$ mesons, such a study has
been performed in Ref.~\cite{Geng:2010vw}, and it was shown that indeed recoil corrections are sizable.\footnote{See Ref.~\cite{Liu:2011mi} for
a related discussion on the scattering lengths of the pseudoscalar mesons off the heavy-light vector mesons.} In Refs.~\cite{Geng:2010df,Altenbuchinger:2011qn}, covariant chiral perturbation theory (ChPT), supplemented with the extended-on-mass-shell (EOMS) scheme, was applied to study the decay constants of the $D(D^*)$/$B(B^*)$ mesons. It was shown that the covariant ChPT converges faster than its nonrelativistic (heavy-meson) counterpart.
These findings can, to some extent, be deemed
as repercussions of the one-baryon sector. For instance, it has been shown that the EOMS formulation of the baryon ChPT
is capable of  better describing three-flavor observables and their light-quark mass evolutions than its nonrelativistic (heavy-baryon) counterpart, see, e.g., Refs.~\cite{Geng:2008mf,MartinCamalich:2010fp,Geng:2013xn} and references cited therein.

In the present work we study  the interactions of the heavy-light mesons ($D$, $D^*$, $B$, $B^*$ and their strange counterparts) with Nambu-Goldstone bosons (the octet of the lightest pseudoscalar mesons) in covariant ChPT and its unitary version. We calculate the interaction potentials up to next-to-leading order (NLO) and perform an iteration of these potentials  to all orders using the Bethe-Salpeter equation.  It was pointed out that in the covariant calculation of the loop function appearing in the Bethe-Salpeter equation, one loses the heavy-quark spin and flavor symmetry~\cite{Cleven:2010aw}. We study this problem in detail and propose a new renormalization scheme, similar in spirit to the EOMS scheme widely used in the one-baryon sector~\cite{  Gegelia:1999gf,Fuchs:2003qc,Geng:2013xn} and also used in Refs.~\cite{Geng:2010vw,Geng:2010df,Altenbuchinger:2011qn},
to recover heavy-quark spin and flavor symmetry up to $1/M_\mathrm{HL}$ corrections, where $M_\mathrm{HL}$ is a generic heavy-light meson mass.  We apply our
approach to describe the most recent fully dynamical LQCD simulations for the scattering lengths of Nambu-Goldstone bosons off the $D$ mesons ~\cite{Liu:2012zya} and
 fix the relevant low-energy and subtraction constants.~\footnote{It should be noted that recently the $D\pi$, $D^*\pi$, and  $DK$ scattering
 lengths have also been calculated on the lattice using both quark-antiquark and meson-meson interpolating fields~\cite{Mohler:2012na,Mohler:2013rwa} and
 the $D^*_{s0}(2317)$ is found to be a bound state in the $DK$ channel~\cite{Mohler:2013rwa}.} We then solve the corresponding Bethe-Salpeter equations and search for poles in the complex energy plane,
identified as dynamically generated states. We show that a number of $0^+$ and $1^+$ states emerge naturally, including the $D^*_{s0}(2317)$, the $D_{s1}(2460)$ and their bottom-quark counterparts. \footnote{A similar strategy was adopted in Refs.~\cite{Liu:2012zya,Wang:2012bu}, but both studies are limited to  the $0^+$ charm sector, and in addition Ref.~\cite{Wang:2012bu} studied the preliminary LQCD results of Ref.~\cite{Liu:2008rza}.}

This article is organized as follows. In Sec. II, the relevant terms of the effective chiral Lagrangian are summarized and the driving potentials up to NLO are constructed. In Sec. III
we propose a new renormalization scheme to be used in the Bethe-Salpeter equation, which manifestly satisfies the chiral power counting rules and heavy-quark spin and flavor symmetries. We discuss the advantage of this scheme in comparison with others widely used in unitary ChPT.
In Sec. IV, we apply both the unitary heavy-meson and covariant formulations of ChPT to fit the LQCD data and make predictions for the existence of a number of dynamically generated resonances in both the charm and the bottom sectors.  A short summary is given in Sec. V.

\section{Theoretical Framework}
\subsection{Chiral Lagrangian up to next-to-leading order}

Introducing the chiral effective Lagrangians in the present context, one first has to  specify a power counting rule.  In the present work, the  Nambu-Goldstone boson (NGB)  masses $m_\phi$ and the field gradients $\partial_\mu\phi$
are counted as $\mathcal{O}(p )$ as usual, where $\phi$ denotes a NGB boson of the pseudoscalar octet.  For $D$ mesons, the triplets are $P=(D^0,D^+,D_s^+)$ and $P^*_\mu=(D^{*0},D^{*+},D_s^{*+})_\mu$, and for $\bar B$ mesons, they are $P=(B^-,\bar B^0,\bar B^0_s)$ and $P^*_\mu=(B^{*-},\bar B^{*0},\bar B^{0*}_s)_\mu$. Their field gradients $\partial_\mu P$ and $\partial_\nu P^*_\mu$ and masses $m_P$ and $m_{P^*}$ are counted as $\mathcal{O}(1)$. The NGB propagator $\frac{i}{q^2-m_\phi^2}$ is counted as $\mathcal{O}(p^{-2})$, while the
heavy-light pseudoscalar and vector meson propagators $\frac{i}{q^2-m_P^2}$ and $\frac{i}{q^2-m_{P^*}^2}(-g^{\mu\nu}+\frac{q^\mu q^\nu}{m_{P^*}^2})$ are counted as $\mathcal{O}(p^{-1})$. The chiral order of the propagators of the heavy-light
mesons can be understood as follows. As in standard heavy-meson (HM) ChPT, one can write the momentum $q$ as a sum of a large component and a residual small component, i.e., $q=m_P\nu+ k$, where $\nu$ is the velocity of the heavy-light meson and $k$ is the small residual component counted as $\mathcal{O}(p)$. Therefore, the heavy-light pseudoscalar  meson propagator becomes $\frac{i}{2 m_P \nu\cdot k + k^2}\approx\frac{i}{2 m_P \nu\cdot k }$, which is counted as  $\mathcal{O}(p^{-1})$. The same is true for the heavy-light vector meson propagator.

The leading order covariant chiral Lagrangian describing the interactions of the NGBs with the heavy-light pseudoscalar and vector mesons
has the following form:
\begin{eqnarray}
\label{LOLag}
\mathcal{L}^{(1)}&=&\langle\mathcal{D}_\mu P\mathcal{D}^\mu P^\dagger\rangle-m_{P}^2\langle PP^\dagger\rangle-\langle\mathcal{D}_\mu P^{*\nu}\mathcal{D}^\mu P^{*\dagger}_\nu\rangle+m_{P^*}^2\langle P^{*\nu}P^{*\dagger}_\nu\rangle\nonumber\\
&&+i\tilde g_{PP^*\phi}\langle P^*_\mu u^\mu P^\dagger-Pu^\mu P^{*\dagger}_\mu \rangle+\frac{g_{P^*P^*\phi}}{2}\langle (P^*_\mu u_\alpha \partial_\beta P_\nu^{*\dagger}-\partial_\beta P_\mu^* u_\alpha P^{*\dagger}_\nu)\epsilon^{\mu\nu\alpha\beta}\rangle\,,
%\nonumber
\end{eqnarray}
where $m_{P}$ and $m_{P^*}$ are the $P$ and $P^*$ masses in the chiral limit, respectively, and $\langle \ldots \rangle$ denotes trace in the $u$, $d$, and $s$ flavor space. The coupling constant $\tilde g_{PP^*\phi}$ has mass dimension $1$, whereas $g_{P^*P^*\phi}$ is dimensionless. The axial current is defined as
$u_\mu=i(\xi^\dagger\partial_\mu \xi-\xi\partial_\mu \xi^\dagger)$ and the chiral covariant derivative is
\begin{equation}
\mathcal D_\mu P_a=\partial_\mu P_a-\Gamma_\mu^{ba}P_b\,,\qquad \mathcal D^\mu P^\dagger_a=\partial^\mu P^\dagger_a+\Gamma^\mu_{ab}P^\dagger_b\,
\end{equation}
with the vector current $\Gamma_\mu=\frac{1}{2}(\xi^\dagger\partial_\mu \xi+\xi\partial_\mu \xi^\dagger)$. In these equations, $\xi^2=\Sigma=\exp(i\Phi/f_0)$ with $f_0$ being
 the NGB decay constant in the chiral limit and $\Phi$ collecting
the octet of NGB fields:
\begin{equation}
\Phi=\sqrt{2}\left(
\begin{array}{ccc}
\frac{\pi^0}{\sqrt{2}}+\frac{\eta}{\sqrt{6}} & \pi^+ & K^+\\
\pi^- & -\frac{\pi^0}{\sqrt{2}}+\frac{\eta}{\sqrt{6}} & K^0\\
K^- & \bar{K}^0 & -\frac{2}{\sqrt{6}}\eta
\end{array}
\right).
\end{equation}

The coupling $\tilde g_{DD^*\phi}$ is known empirically. It can be determined from the decay width $\Gamma_{D^{*+}}=(96\pm22)\,\textrm{keV}$ together with the branching ratio $BR_{D^{*+}\rightarrow D^0\pi^+}=(67.7\pm0.5)\%$~\cite{Beringer:1900zz}. At tree level, $\Gamma_{D^{*+}\rightarrow D^0\pi^+}=\frac{1}{12\pi}\frac{\tilde g_{DD^*\phi}^2}{f_0^2}\frac{|q_\pi|^3}{M^2_{D^{*+}}}$, which gives $\tilde g_{DD^*\phi}=(1177\pm137)\,\textrm{MeV}$.   The coupling $ g_{D^*D^*\phi}$ can be related to $\tilde{g}_{DD^*\phi}$ through the heavy-quark spin symmetry, i.e., $g_{D^*D^*\phi} M_{D^*}=\tilde{g}_{DD^*\phi}$, keeping in mind that there could be sizable deviations of higher order in $1/m_D$. The couplings $g_{BB^*\phi}$ and $g_{B^*B^*\phi}$ can be related to
their $D$ counterparts through heavy-quark flavor symmetry.

In a similar way, one can construct the covariant NLO terms of the effective Lagrangian:
\begin{eqnarray}
\label{NLOLag}
\mathcal{L}^{(2)}&=&-2[c_0\langle P P^\dagger\rangle\langle \chi_+\rangle-c_1\langle P \chi_+ P^\dagger\rangle-c_2\langle P P^\dagger\rangle\langle u^\mu u_\mu\rangle-c_3\langle P u^\mu u_\mu P^\dagger\rangle\nonumber\\&&+\frac{c_4}{m_P^2}\langle \mathcal D_\mu P\mathcal D_\nu P^\dagger\rangle\langle\{u^\mu,u^\nu\}\rangle+\frac{c_5}{m_P^2}\langle \mathcal D_\mu P\{u^\mu,u^\nu\} \mathcal D_\nu P^\dagger\rangle+\frac{c_6}{m_P^2}\langle \mathcal D_\mu P[u^\mu,u^\nu]\mathcal D_\nu P^\dagger\rangle]\nonumber\\
&&+2[\tilde c_0\langle P^*_\mu P^{*\mu\dagger}\rangle\langle \chi_+\rangle-\tilde c_1\langle P^*_\mu \chi_+ P^{*\mu\dagger}\rangle- \tilde c_2\langle P^*_\mu P^{*\mu\dagger}\rangle\langle u^\mu u_\mu\rangle-\tilde c_3\langle P^*_\nu u^\mu u_\mu P^{*\nu\dagger}\rangle\nonumber\\&&+\frac{\tilde c_4}{m_{P^*}^2}\langle \mathcal D_\mu P^*_\alpha\mathcal D_\nu P^{*\alpha\dagger}\rangle\langle\{u^\mu,u^\nu\}\rangle+\frac{\tilde c_5}{m_{P^*}^2}\langle \mathcal D_\mu P^*_\alpha\{u^\mu,u^\nu\} \mathcal D_\nu P^{*\alpha\dagger}\rangle\nonumber\\&&+\frac{\tilde c_6}{m_{P^*}^2}\langle \mathcal D_\mu P_\alpha^{*}[u^\mu,u^\nu]\mathcal D_\nu P^{*\alpha\dagger}\rangle]\,,\label{NLOLag}
\end{eqnarray}
where $\chi_+=\xi^\dagger \mathcal{M} \xi^\dagger + \xi\mathcal{M}\xi$ with $\mathcal{M}= \mathrm{diag}(m_\pi^2,m_\pi^2,2 m_K^2-m_\pi^2)$.

In the infinite heavy-quark mass limit, one has $c_i=\tilde c_i$ for $i=0,\ldots6$ and $m_P=m_{P^*}$. For the numerical results presented in this work, we have fixed the $m_P$ and $m_{P^*}$ appearing in
Eq.~(\ref{NLOLag}), which are needed to make  the $c_4$, $c_5$, and $c_6$  low-energy constants (LECs) dimensionless, to the following values:
$m_D=m_{D^*}=\mathring{m}_D$ and $m_B=m_{B^*}=\mathring{m}_B$ (see Table \ref{table:par}), where $\mathring{m}_D$ ($\mathring{m}_B$) is the SU(3) average of strange and non-strange D(B) and $D^*(B^*)$ masses. Such a choice is taken in order to avoid introducing SU(3) breaking corrections to the LECs by hand in the covariant framework.
As a first estimate of the size of spin or flavor symmetry-breaking effects, one can determine the constants $c_1$ and $\tilde c_1$ from the masses of strange and nonstrange $D$ and $D^*$ mesons. At the NLO chiral order, the masses of the $D$, $D_s$, $D^*$ and $D^*_s$ mesons are given by
\begin{eqnarray}
M_D^2&=&m_D^2+4 c_0(m_\pi^2+2m_K^2)-4c_1 m_\pi^2\,,\label{MassForm1}\\
M_{D_s}^2&=&m_D^2+4 c_0(m_\pi^2+2m_K^2)+4c_1 (m_\pi^2-2m_K^2)\,,\label{MassForm2}\\
M_{D^*}^2&=&m_{D^*}^2+4 \tilde c_0(m_\pi^2+2m_K^2)-4\tilde c_1 m_\pi^2\,,\label{MassForm3}\\
M_{D^*_s}^2&=&m_{D^*}^2+4 \tilde c_0(m_\pi^2+2m_K^2)+4\tilde c_1 (m_\pi^2-2m_K^2)\,,
\label{MassForm4}
\end{eqnarray}
where the $D (D^*)$ meson mass in the chiral limit is denoted as $m_{D}$ $(m_{D^*})$.
Inserting the physical masses listed in Table \ref{table:par} leads to $c_1=-0.214$ and $\tilde c_1=-0.236$.
Repeating the same argument for the $\bar{B}$ mesons, we obtain $c_1(B)=-0.513$ and $\tilde{c}_1(B)=-0.534$.
The heavy-quark  flavor symmetry dictates that  $c_1(\tilde{c}_1)/M_\mathrm{HL}$=const. Using an SU(3)-averaged mass for $M_\mathrm{HL}$ for each sector,
we find $c_1/\bar{M}_D=-0.113$ GeV$^{-1}$, $\tilde{c}_1 /\bar{M}_{D^*}=-0.116$ GeV$^{-1}$, $c_1(B)/\bar{M}_B=-0.097$ GeV$^{-1}$, and $\tilde{c}_1(B)/\bar{M}_{B^*}=-0.100$ GeV$^{-1}$. These numbers provide a hint about
the expected order of magnitude for the breaking of heavy-quark spin and flavor symmetry: about $3\%$ between $D$ vs $D^*$ and $B$ vs $B^*$, whereas it amounts to about $16\%$ between $D$ vs $B$ and $D^*$ vs $B^*$.

\subsection{Chiral potentials}
\begin{figure}[t]
\includegraphics[width=1\textwidth]{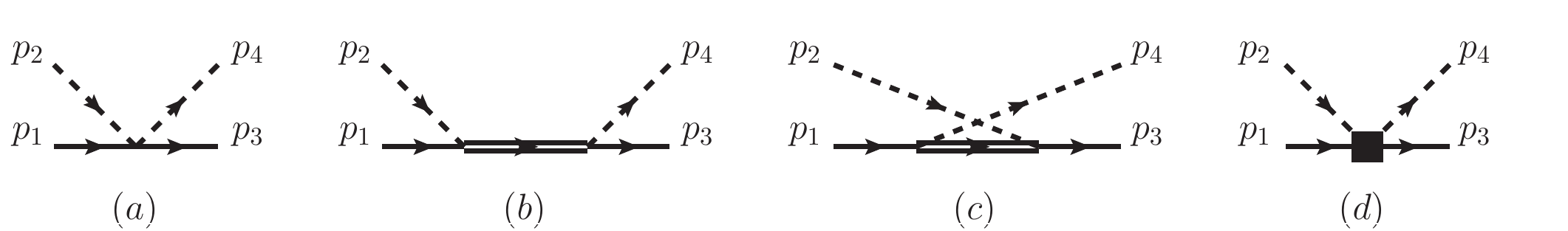}
\caption{Feynman diagrams contributing to $P \phi\rightarrow P\phi$ at LO (a)-(c) and NLO (d) chiral order. The pseudoscalar ($P$) mesons are represented by solid lines, the vector ($P^*$) mesons by double lines, and the Nambu-Goldstone bosons by dashed lines.}
\label{LOFeynman}
\end{figure}

\begin{figure}
\includegraphics[width=0.8\textwidth]{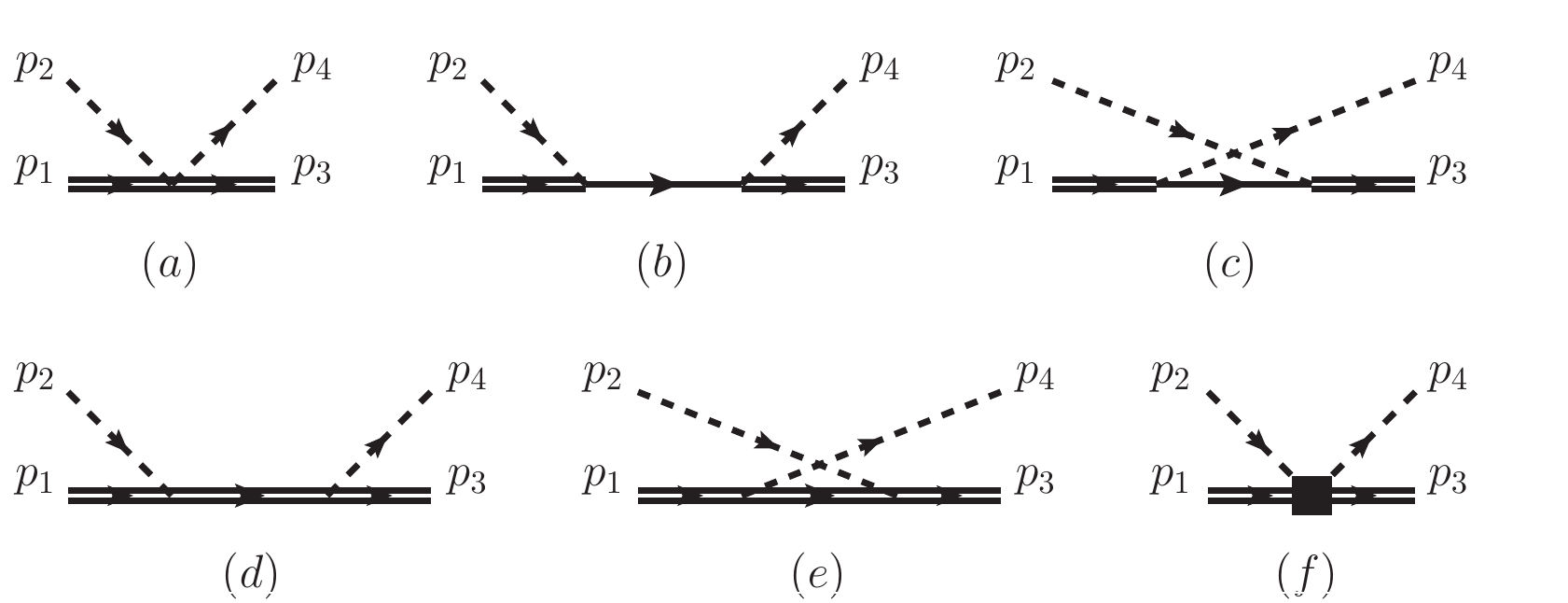}
\caption{Feynman diagrams contributing to $P^*\phi\rightarrow P^*\phi$ at LO (a)-(e) and NLO (f) chiral order.
The pseudoscalar ($P$) mesons are represented by solid lines, the vector ($P^*$) mesons by double lines, and the Nambu-Goldstone bosons by dashed lines. }
\label{VectorLOFeynman}
\end{figure}

In this section we derive the chiral potentials contributing to $P^{(*)}\phi\rightarrow P^{(*)}\phi$ scattering up to NLO in both the covariant  and the heavy-meson formulations. The corresponding Feynman diagrams are shown in Figs.~\ref{LOFeynman} and
~\ref{VectorLOFeynman}.

\subsubsection{$J^P=0^+$ potential for $P\phi\rightarrow P\phi$}
For the processes $P\phi\rightarrow P\phi$, the leading order (LO) potential can be written as
\begin{equation}
\mathcal{V}_\mathrm{LO}=\mathcal{V}_\mathrm{WT}+\mathcal{V}_\mathrm{s-Ex}+\mathcal{V}_\mathrm{u-Ex},
\end{equation}
where $\mathcal{V}_\mathrm{WT}$, $\mathcal{V}_\mathrm{s-Ex}$ and $\mathcal{V}_\mathrm{u-Ex}$ are
the Weinberg-Tomozawa term and the $s$- and $t$-channel exchange contributions, respectively.
The Weinberg-Tomozawa term $\mathcal{V}_\mathrm{WT}$ has the following form:
\begin{eqnarray}
\mathcal V_{\rm{WT}}(P(p_1)\phi(p_2)\rightarrow P(p_3)\phi(p_4))&=&%-
\frac{1}{4 f_0^2}\mathcal C_{\text{LO}}
   \left(s-u\right)\,,
   \label{LOPot}
   \end{eqnarray}
with the Mandelstam variables $s=(p_1+p_2)^2=(p_3+p_4)^2$ and
$u=(p_1-p_4)^2=(p_3-p_2)^2$. The coefficients $\mathcal C_{\rm{LO}}$ for different strangeness and isospin combinations $(S,I)$ are listed in Table \ref{Coefficients}.
The $s/u$-channel exchange terms $\mathcal{V}_\mathrm{s-Ex}$ and $\mathcal{V}_\mathrm{u-Ex}$ are suppressed by $1/M_\mathrm{HL}$ compared to the Weinberg-Tomozawa term. At threshold
they are in fact of second chiral order and can be absorbed into the available $\mathcal{O}(p^2)$ LECs. In addition, we have checked numerically that they play a negligible role in the present study,  and hence we neglect their contributions in the following. The same statements hold for the $u$-channel exchange diagrams in the $P^*\phi\rightarrow P^*\phi$ process. The $s$-channel diagrams in the $P^*\phi\rightarrow P^*\phi$ process do not contribute to $S$-wave interactions.

\begin{table*}[t]
      \renewcommand{\arraystretch}{1.6}
     \setlength{\tabcolsep}{0.55cm}
     \centering
     \caption{\label{table:par}Numerical values of isospin-averaged masses and the pion decay constant $f_0$ (in units of MeV)~\cite{Beringer:1900zz}. The eta meson mass is calculated using the Gell-Mann-Okubo mass relation: $m_\eta^2=(4 m_K^2-m_\pi^2)/3$.}
     \begin{tabular}{ccccccccc}
     \hline\hline
    $\mathring{m}_{D}$  &  $M_{D^*_s}$ & $M_{D^*}$ & $M_{D_s}$   & $M_D$  &  $m_\pi$ & $m_K$  &    $m_\eta$      \\
   $1972.1$    & 2112.3 & 2008.6  & 1968.5  & 1867.2 & 138.0  & 495.6 & 566.7 \\   \hline
    $\mathring{m}_{B}$  &  $M_{B^*_s}$ & $M_{B^*}$ & $M_{B_s}$  & $M_B$     & $f_0$  \\
   $5331.9$    & 5415.4 &   5325.2&   5366.8& 5279.4 & 92.21 \\
 \hline\hline
    \end{tabular} % \par
\end{table*}

The NLO potential has the following form:
 \begin{eqnarray}
\mathcal V_{\rm{NLO}}(P(p_1)\phi(p_2)\rightarrow P(p_3)\phi(p_4))&=&
%+
-\frac{8}{f_0^2} C_{24} \left(c_2\, p_2\cdot p_4-\frac{c_4}{m^2_P}\left( p_1\cdot p_4\; p_2\cdot
   p_3+ p_1\cdot p_2\; p_3\cdot p_4\right)\right)
   \nonumber\\&&%+
    - \frac{4}{f_0^2} \mathcal C_{35}
   \left(c_3\, p_2\cdot p_4-\frac{c_5}{m^2_P} \left(p_1\cdot p_4 \;p_2\cdot
   p_3+p_1\cdot p_2 \;p_3\cdot p_4\right)\right)\nonumber\\
   &&%+
   -\frac{4}{f_0^2} \mathcal C_{6}\,\frac{c_6}{m^2_P} \left(p_1\cdot p_4 \;p_2\cdot
   p_3-p_1\cdot p_2 \;p_3\cdot p_4\right)\nonumber\\
   &&%+
   -\frac{8}{f_0^2}
   \mathcal C_0\,c_0%-
   +\frac{4}{f_0^2} \mathcal C_1\,c_1\, ,
   \label{NLOPot}
\end{eqnarray}
where the coefficients $\mathcal C_i$ can be found in Table \ref{Coefficients}.

The LECs $c_0,...,c_6$ in the $B$ and $D$ meson sectors are related by $c_{i,B}/\mathring{m}_B=c_{i,D}/\mathring{m}_{D}$  up to corrections in $1/\mathring{m}_B(\mathring{m}_D)$, where $\mathring{m}_D$ and $\mathring{m}_B$ are the generic $D$ and $B$ meson masses, respectively, given in Table \ref{table:par}. Since the $P$ and $P^*$ masses
are very close to each other, one can use $c_{i,P^*}=c_{i,P}$,  again up to corrections in $1/\mathring{m}_P$ for $P=B,D$.

In the present case we are only interested in $S$-wave interactions and, therefore, can project the potentials accordingly:
\begin{equation}
\mathcal{V}_{\mathrm{LO}/\mathrm{NLO}}|_\mathrm{s-wave}=\frac{1}{2}\int\limits^{1}_{-1} \mathcal{V}_{\mathrm{LO}/\mathrm{NLO}}\, d\cos(\theta),
\end{equation}
where $\theta$ is the angle between the three-momenta of the initial  and final heavy-light mesons.

It should be pointed out that the terms multiplying $c_6$ vanish at threshold. Furthermore,
they have negligible effects on the dynamical generation of bound or resonant states as long as $c_6$ is of natural size. Therefore we are not going to consider the $c_6$ terms further in the present work.

\subsubsection{$J^P=1^+$ potential for $P^*\phi \rightarrow P^*\phi$}

From the Lagrangian (\ref{LOLag},\ref{NLOLag}),  one can easily compute the corresponding LO and NLO potentials
\begin{eqnarray}
\mathcal V_{\textrm{LO(NLO)}}(P^*(p_1)\phi(p_2)\rightarrow P^*(p_3)\phi(p_4))&=&-\epsilon^*_3\cdot\epsilon_1\,\mathcal V_{\textrm{LO(NLO)}}(P(p_1)\phi(p_2)\rightarrow P(p_3)\phi(p_4)).
\label{VecPot}
\end{eqnarray}
The polarization vectors are treated by turning to a representation in terms of helicity states which allows one to
decompose the potentials into subsectors of good angular momentum and parity.
The potentials thus acquire a matrix structure.\footnote{It should be pointed out that in the infinite heavy-quark limit, one has $\epsilon^*_3\cdot\epsilon_1=-1$,  which leads to
$\mathcal V_{\textrm{LO(NLO)}}(P^*(p_1)\phi(p_2)\rightarrow P^*(p_3)\phi(p_4))=\mathcal V_{\textrm{LO(NLO)}}(P(p_1)\phi(p_2)\rightarrow P(p_3)\phi(p_4))$ (see also, e.g., Ref.~\cite{Abreu:2011ic}). } Such a procedure is explained in detail in Ref.~\cite{Lutz:2003fm}. In the present work, we are interested in
the $J^P=1^+$ sector and the potential matrix becomes
\begin{equation}
\hat{V} \equiv\left(\begin{array}{cc}\langle 1^+|\mathcal V^{J=1}|1^+\rangle&\langle 1^+|\mathcal V^{J=1}|0\rangle\\\langle 0|\mathcal V^{J=1}|1^+\rangle&\langle 0|\mathcal V^{J=1}|0\rangle\end{array}\right)\,,
\label{VecPot}
\end{equation}
where the matrix elements can be straightforwardly computed following the procedure outlined in the Appendix of Ref.~\cite{Lutz:2003fm}. In order to construct projectors free of kinematic singularities, the bare helicity states have been rotated, resulting in the following normalizations \cite{Lutz:2003fm}:
\begin{equation}
N=\left(\begin{array}{cc}\langle 1^+|1^+\rangle&\langle 1^+|0\rangle\\\langle 0|1^+\rangle&\langle 0|0\rangle\end{array}\right)=
\left(\begin{array}{cc} \frac{3}{2}+\frac{p^2_\mathrm{cm}}{2M^2}&\frac{p^2_\mathrm{cm}\sqrt{M^2+p^2_\mathrm{cm}}}{\sqrt{2}M^2}\\
\frac{p^2_\mathrm{cm}\sqrt{M^2+p^2_\mathrm{cm}}}{\sqrt{2}M^2}&\frac{p^4_\mathrm{cm}}{M^2}\end{array}\right),
\end{equation}
where $p_\mathrm{cm}$ is the center of mass three-momentum of the interacting pair.
We have checked  numerically that the matrix elements involving the helicity state $|0\rangle$ play a negligible role in our present study.\footnote{This can be naively understood as follows. In
the non-relativistic limit, the $|1\rangle$ state is built from S-wave interactions while the $|0\rangle$ state is built from D-wave interactions. Since we are not far away from threshold, the D-wave interactions and the S-D transitions seem to be small, as suggested by
the actual numerical analysis. } Therefore we only keep the  $\hat{V}_{11}$ component of the potential,
which coincides with the approach of Ref.~\cite{Roca:2005nm}.
\begin{table}
\caption{Coefficients of the LO and NLO potentials for $D\phi \rightarrow D\phi$ [Eqs.(\ref{LOPot}),(\ref{NLOPot}),(\ref{HMLOPot}),(\ref{HMNLOPot})].
The coefficients for $D^*\phi\rightarrow D^*\phi$ or $B^{(*)}\phi\rightarrow B^{(*)}\phi$ can
be obtained by replacing  $D/D_s$ with $D^*/D^*_s$ or $B^{(*)}/B^{(*)}_s$.}
\label{Coefficients}
\begin{equation}
\begin{array}{cc|cccccc}
\hline\hline
 (S,I) & \text{Channel} & \mathcal C_{\text{LO}} & \mathcal C_0 &\mathcal C_1 &\mathcal C_{24}
   &\mathcal C_{35} &\mathcal C_6 \\
   \hline
   \hline
 \text{(2,1/2)} & D_sK\to D_sK & 1 & m_K{}^2 & m_K{}^2 & 1 & 1 & -1
   \\
   \hline
 \text{(1,1)} & DK\to DK & 0 & m_K{}^2 & 0 & 1 & 0 &
   0 \\
 \text{} & D_s\pi \to D_s\pi  & 0 & m_{\pi }{}^2 & 0 & 1 & 0 & 0 \\
 \text{} & DK\to D_s\pi  & 1 & 0 & \frac{1}{2}
   \left(m_K{}^2+m_{\pi }{}^2\right) & 0 & 1 & -1 \\
   \hline
 \text{(1,0)} & DK\to DK & -2 & m_K{}^2 & 2 m_K{}^2 &
   1 & 2 & 2 \\
 \text{} & D_s\eta \to D_s\eta  & 0 & \frac{1}{3} \left(4
   m_K{}^2-m_{\pi }{}^2\right) & \frac{4}{3} \left(2 m_K{}^2-m_{\pi
   }{}^2\right) & 1 & \frac{4}{3} & 0 \\
 \text{} & DK\to D_s\eta  & -\sqrt{3} & 0 & \frac{5
   m_K{}^2-3 m_{\pi }{}^2}{2 \sqrt{3}} & 0 & \frac{1}{\sqrt{3}} &
   \sqrt{3} \\
   \hline
 \text{(0,3/2)} &D\pi\to D\pi  & 1 & m_{\pi
   }{}^2 & m_{\pi }{}^2 & 1 & 1 & -1 \\
   \hline
 \text{(0,1/2)} & D\pi \to D\pi  & -2 & m_{\pi
   }{}^2 & m_{\pi }{}^2 & 1 & 1 & 2 \\
 \text{} & D\eta \to D\eta  & 0 & \frac{1}{3}
   \left(4 m_K{}^2-m_{\pi }{}^2\right) & \frac{m_{\pi }{}^2}{3} & 1
   & \frac{1}{3} & 0 \\
 \text{} & D_s\bar{K}\to D_s\bar{K} & -1 & m_K{}^2 & m_K{}^2 & 1 &
   1 & 1 \\
 \text{} & D\pi \to D\eta  & 0 & 0 & -m_{\pi
   }{}^2 & 0 & -1 & 0 \\
 \text{} & D\pi \to D_s\bar{K} & \sqrt{\frac{3}{2}} & 0 &
   -\frac{1}{2} \sqrt{\frac{3}{2}} \left(m_K{}^2+m_{\pi
   }{}^2\right) & 0 & -\sqrt{\frac{3}{2}} & -\sqrt{\frac{3}{2}} \\
 \text{} & D\eta \to D_s\bar{K} & -\sqrt{\frac{3}{2}} & 0
   & \frac{3 m_{\pi }{}^2-5 m_K{}^2}{2 \sqrt{6}} & 0 &
   -\frac{1}{\sqrt{6}} & \sqrt{\frac{3}{2}} \\
   \hline
 \text{(-1,1)} & D\bar{K}\to D\bar{K} & 1 & m_K{}^2 & m_K{}^2 & 1 &
   1 & -1 \\
   \hline
 \text{(-1,0)} & D\bar{K}\to D\bar{K} & -1 & m_K{}^2 & -m_K{}^2 & 1
   & -1 & 1\\
   \hline
\end{array}
\nonumber
\end{equation}
\end{table}

\subsubsection{Heavy-meson ChPT}
In the heavy-meson (HM) ChPT at LO, the Weinberg-Tomozawa potential $\mathcal{V}_\mathrm{WT}$ reduces to
\begin{equation}\label{HMLOPot}
\mathcal{V}_\mathrm{WT}=\frac{\mathring{m}}{2 f_0^2}(E_2 + E_4) \,\mathcal{C}_\mathrm{LO}
\end{equation}
with
$\mathring{m}$ given in Table \ref{table:par}. At NLO, with the on-shell approximation and for $S$-wave interactions, effectively only four of the six low-energy constants contribute, i.e.,
 \begin{eqnarray}
\mathcal V_{\rm{NLO}}=
-\frac{8}{f_0^2} C_{24} \,c_{24}\, E_2 E_4
    - \frac{4}{f_0^2} \mathcal C_{35}\,
   c_{35}\, E_2 E_4
   -\frac{8}{f_0^2}
   \mathcal C_0\,c_0%-
   +\frac{4}{f_0^2} \mathcal C_1\,c_1\, ,
   \label{HMNLOPot}
\end{eqnarray}
where $c_{24}=c_2-2c_4$ and $c_{35}=c_3-2c_5$ (see, e.g., Ref.~\cite{Cleven:2010aw}).

\section{Bethe-Salpeter equation and renormalization scheme motivated by heavy-quark symmetry}

It is well known that perturbation theory at any finite order cannot generate bound states or resonances. One way to proceed is to perform an infinite summation of a leading subclass of diagrams to all orders using the Bethe-Salpeter (or Lippmann-Schwinger) equation. In combination with coupled-channels dynamics,
this approach has turned out to be quite successful in describing a multitude of low-energy strong-interaction phenomena (see, e.g., Refs.~\cite{Kaiser:1995eg,Dobado:1996ps,Oller:1997ti,Oset:1997it,Oller:1998hw,Kaiser:1998fi,Oller:1998zr,Oller:2000fj,Lutz:2001yb} for early references and
Refs.~\cite{Garcia-Recio:2013uva,Ozpineci:2013zas} for some recent applications). To simplify the calculations, the so-called on-shell approximation~\cite{ Oller:1997ti,Oset:1997it} is often
introduced, with the argument that the off-shell effects are relegated to higher orders. See Ref.~\cite{Altenbuchinger:2013gaa} for a comparison of the on-shell approximation and
the full results with off-shell effects taken into account. The results presented there show that in the $D\phi$ sector, the on-shell and off-shell approaches yield similar results, indicating that to a large extent the off-shell effects can be absorbed into the local counter terms. Since the only scale of relevance in different sectors is the heavy-light meson mass, it is reasonable to assume that these LECs in different sectors, related to each other by the $1/M_{HL}$ relation,  should be able to take into account the off-shell effects, without spoiling the heavy-quark spin/flavor symmetry in  any dramatic way. Therefore, we adopt the on-shell approximation in the present work.

Schematically, the Bethe-Salpeter equation can be
written as
\begin{equation}
T=V+VGT,
\end{equation}
where $V$ is the potential and $G$ is a loop function defined in the following way
\begin{equation}
G(s,M^2,m^2)\equiv i\int \frac{d^n q}{(2\pi)^n}\frac{1}{[(P-q)^2-m^2+i\epsilon][q^2-M^2+i\epsilon]},
\end{equation}
where $P=(\sqrt{s},0,0,0)$ is the total momentum of the two particles. $M$ and $m$ are the masses of the heavy-light meson and of the NGB, respectively,
in the two-particle intermediate state. According to the power counting rule specified in Sec. II, the loop function $G$ counts as $\mathcal{O}( p )$. An explicit evaluation in $n = 4$ dimensions with the modified minimal subtraction scheme yields
\begin{eqnarray}
G_{\overline{\mathrm{MS}}}(s,M^2,m^2)&=&\frac{1}{16 \pi^2 }\Big\{ \frac{m^2-M^2+s}{2s} \log \left(\frac{m^2}{M^2}\right)\nonumber\\
&&
-\frac{q}{\sqrt{s}} \left\{\log[2 q \sqrt{s}+m^2-M^2-s]+\log[2 q \sqrt{s}-m^2+M^2-s]\right.\nonumber\\
&&\left. - \log[2 q \sqrt{s}+m^2-M^2+s]-\log[2 q \sqrt{s}-m^2+M^2+s]\right\}\nonumber\\
&&+ \underline{\left(\log\left(\frac{M^2}{\mu^2}\right) -2\right)}\Big\}\,,
   \label{LoopF}
\end{eqnarray}
where  $q=\frac{\sqrt{(s-(m+M)^2)(s-(m-M)^2)}}{2\sqrt{s}}$ is the center of mass (three-)momentum.
It is easily seen that the underlined term in the loop function (\ref{LoopF}) breaks  the chiral power counting. In addition, the heavy-quark flavor symmetry and, to a less extent, the heavy-quark spin symmetry are also broken in the covariant loop function, as noticed in Ref.~\cite{Cleven:2010aw}.  To take into account nonperturbative physics, the usual practice in the unitary ChPT (UChPT) is to replace the underlined term $-2$ by the so-called subtraction constant $a(\mu)$, which we will refer to as the $\overline{\mathrm{MS}}$ scheme.

In the following we propose a renormalization scheme that restores the chiral power counting and ensures that the loop function $G$ has a well-defined behavior in the $M\rightarrow \infty$ limit.  To achieve this, we turn to the HM ChPT, where the loop function takes the following form (see, e.g., Refs.~\cite{Scherer:2002tk,Cleven:2010aw})
\begin{eqnarray}
G_{\textrm{HM}}(s,M^2,m^2)
&=&\frac{1}{16\pi^2  \mathring{M} }\Big\{2\sqrt{
\Delta_\textrm{HM} ^2-m^2} \left(\mathrm{arccosh}\left(\frac{\Delta_\mathrm{HM}}{m}\right)-\pi i\right)+
 \Delta_\textrm{HM}\left(  \log\left(\frac{m^2}{\mu^2}\right) +a \right) \Big\}\,,\nonumber\\
\label{HMChPT}
\end{eqnarray}
where $\mathring{M}$ is the chiral limit value of the heavy-light meson mass appearing in the loop  and $\Delta_\mathrm{HM}=\sqrt{s}-M$.
Comparing $G_\mathrm{HM}$ with the loop function  of Eq.~(\ref{LoopF}) expanded up to $1/\mathring{M}$
\begin{eqnarray}
& & G(s,M^2,m^2)=\underline{\frac{1}{16\pi ^2}
\left(\log \left(\frac{\mathring{M} ^2}{\mu^2}\right)-2\right)} \nonumber\\
& & +\frac{1}{16\pi^2  \mathring{M}}\Big\{2\sqrt{
\Delta_{\textrm{HM}} ^2-m^2}\left(\mathrm{arccosh}\left(\frac{\Delta_\mathrm{HM}}{m}\right)-\pi i\right)+\Delta_{\textrm{HM}}  \log\left(\frac{m^2}{\mathring{M}^2}\right)\Big\},
\label{1overMExp}
\end{eqnarray}
 one is tempted to introduce the following renormalization scheme:
\begin{eqnarray}\label{GHQS}
G_\mathrm{HQS}(s,M^2,m^2)&\equiv&
G(s,M^2,m^2)  \nonumber\\
&-&\frac{1}{16\pi ^2}\left(\log\left(\frac{\mathring{M}^2}{\mu^2}\right)-2\right)
+\frac{m_\mathrm{sub}}{16\pi^2 \mathring{M}} \left( \log\left(\frac{\mathring{M}^2}{\mu^2}\right) + a\right),
\end{eqnarray}
where $m_\mathrm{sub}=m$. From now on, we will refer to this loop function as the heavy-quark symmetry (HQS) inspired loop function and the covariant UChPT with the HQS inspired loop function also as the HQS UChPT.   It should be noted that in Eq.~(\ref{GHQS}) we have chosen to renormalize the loop function at the threshold of $\sqrt{s}=M+m$, where $\Delta_\mathrm{HM}=m(m_\mathrm{sub})$.   The renormalized loop function $G_\mathrm{HQS}$  satisfies the chiral power counting and also exhibits a well-defined behavior in the $M\rightarrow\infty$ limit.\footnote{
It is clear that if we drop the nonperturbative term $\frac{m }{16\pi^2 \mathring{M}}\left( \log\left(\frac{\mathring{M}^2}{\mu^2}\right)+a \right)$, our proposed renormalization scheme is in spirit similar to the EOMS scheme widely used in the one-baryon sector to remove the power-counting-breaking terms.}
At fixed $\mathring{M}$ and $m_\mathrm{sub}$ the ansatz we propose
is equivalent to the $\overline{\mathrm{MS}}$ approach widely used in UChPT, but it has the advantage of manifestly satisfying the (approximate) heavy-quark spin and flavor symmetries.

In our study of the scattering lengths of NGB bosons off the $D$ mesons, the subtraction constant $a$ can in principle vary from channel to channel, depending on the intermediate NGB.
A reasonable  alternative is to  use for $m_\mathrm{sub}$ an SU(3) average NGB mass, e.g., $m_\mathrm{sub}=(3 m_\pi + 4 m_K + m_\eta)/8=0.3704$ GeV, and have a common subtraction constant $a$ for all channels.  A variation of this value from $m_\pi$ to $m_\eta$ can serve as an estimate of uncertainties as one tries to connect physics of the $D$ and $B$ sectors. It should be stressed that using the mass of the intermediate NGB in the subtraction but keeping a common subtraction constant for all channels will  introduce sizable uncontrolled SU(3)-breaking corrections that should be avoided.

\begin{figure}
\includegraphics[angle=270,width=0.8\textwidth]{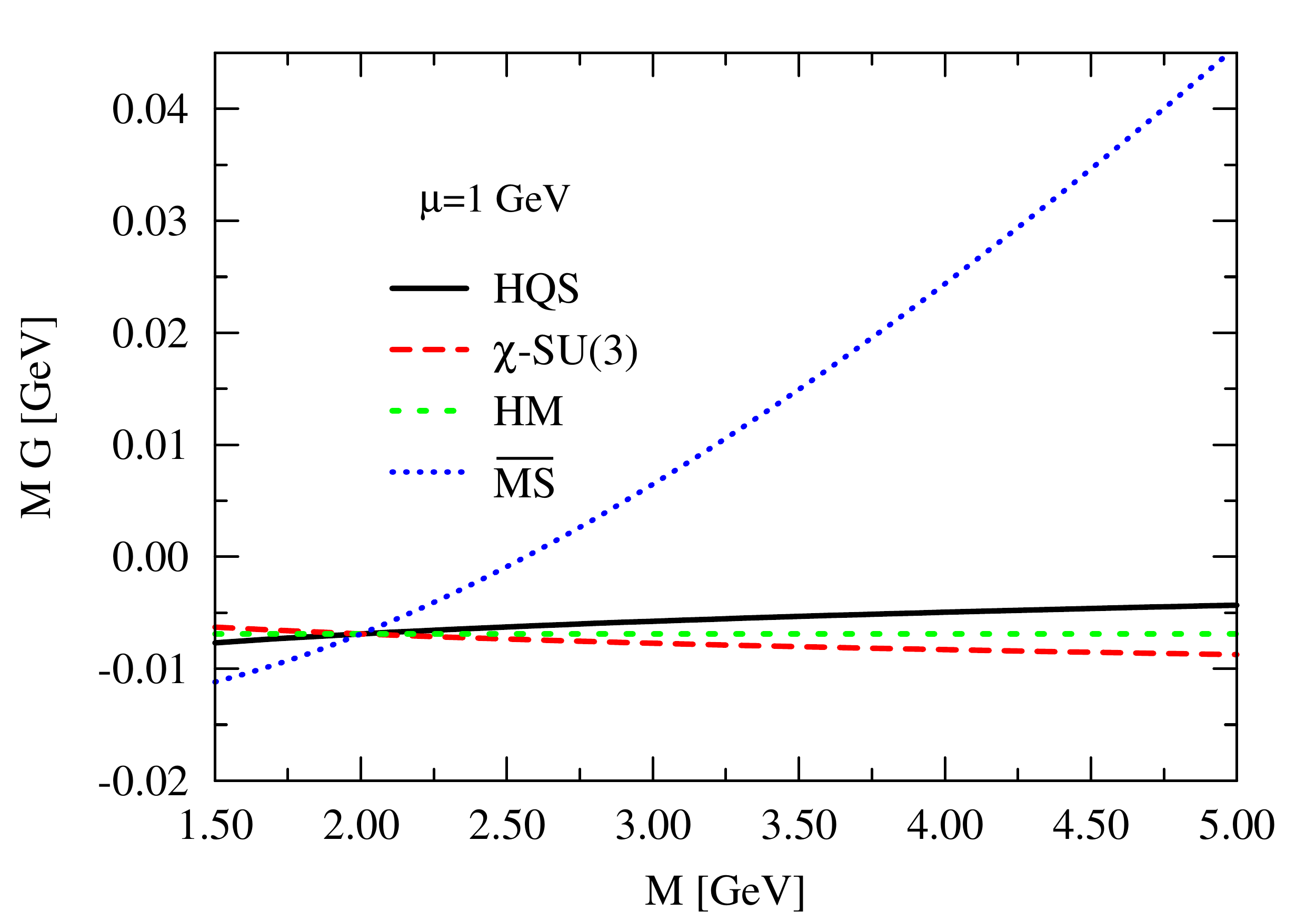}
\caption{Dependence of loop functions (at threshold) on the heavy-light meson mass in different schemes with $\mu=1$ GeV. \label{fig:mdep}}
\end{figure}

In Fig.~\ref{fig:mdep}, we show the dependence of the loop functions  on the heavy-light meson mass $M$, calculated in the HQS, HM and $\overline{\mathrm{MS}}$ schemes with
the renormalization scale $\mu=1$ GeV,~\footnote{From a theoretical point of view, the renormalization scale $\mu$ should be the chiral-symmetry breaking scale, $\Lambda_\chi\approx 4\pi f_0\approx 1.2$ GeV, which can be immediately seen by examining the HM ChPT loop function of Eq.~(\ref{HMChPT}).}  $\mathring{M}=M$,  $m=m_\pi=0.138$ GeV, $\sqrt{s}=M+m$, and $m_\mathrm{sub}=0.3704$ GeV.  For the sake of comparison, we have
plotted the loop function defined in the chiral SU(3) scheme of Ref.~\cite{Lutz:2001yb}, which has the following form:
\begin{equation}
G_{\chi-\mathrm{SU(3)}}=G_{\overline{\mathrm{MS}}}(s,M^2,m^2)-G_{\overline{\mathrm{MS}}} (M^2,M^2,m^2).
\end{equation}
The subtraction constants in
the HM, HQS, and $\overline{\mathrm{MS}}$ schemes are adjusted to reproduce the $G_{\chi-\mathrm{SU(3)}}$ at $M=2$ GeV.
From Eq.~(\ref{HMChPT}) one can see that  $G_\mathrm{HM}$ is
 inversely proportional to $M$
and therefore $M G$ is a constant for the HM loop function. On the other hand,
the $G$ function in the HQS scheme is slightly upward curved while the $G$ function in the $\chi$-SU(3) downward curved.  The
naive $\overline{\mathrm{MS}}$ scheme, on the other hand, changes rapidly with $M$.  It is clear that without readjusting $a$ for different $M$, which could correspond to
either a heavy-light $B$ meson or $D$ meson, heavy-quark flavor symmetry is lost as pointed out in Ref.~\cite{Cleven:2010aw}.

So far, we have concentrated on the $1/M$ scaling of the loop function $G$ in different schemes but have not paid much attention to the chiral series or
SU(3)-breaking effects. In terms of $1/M$ scaling, the HM, HQS, and $\chi$-SU(3) approaches all seem reasonable, as shown in Fig.~\ref{fig:mdep}.
On the other hand, compared to the HM ChPT or the $\chi$-SU(3) approach, the subtraction constant in the HQS scheme has the simplest form consistent
with the chiral power counting and $1/M$ scaling.  We will see in the following section that such a choice seems to play a non-negligible role in describing the light-quark mass dependence of
the scattering lengths of the NGBs off the $D$ mesons.

\section{Results and discussions}
\subsection{Fits to the LQCD data of scattering lengths}

Now we are in a position to study the latest fully dynamical  LQCD data of Ref.~\cite{Liu:2012zya}. Up to NLO,\footnote{It should be noted
that the scattering lengths of the NGBs off the $D$ mesons have been calculated up to N$^3$LO in
both the covariant ChPT~\cite{Geng:2010vw} and HM ChPT~\cite{Liu:2009uz}.} we have six unknown LECs and in the case of the
UChPT also the unknown subtraction constant. As explained in Section 2, the constant $c_1$ can be determined from the mass splitting of the strange and nonstrange $D$ mesons,
which yields $c_1=-0.214$. The constant $c_0$ can be fixed by fitting the NLO mass formulas to the LQCD data of Ref.~\cite{Liu:2012zya}. This yields
$c_0=0.015$.  Therefore, we have four LECs to be determined in the ChPT and five in the UChPT. In our
framework, the scattering lengths of channel $i$ with strangeness $S$ and isospin $I$ are
related to the diagonal $T$-matrix elements $T_{ii}$ via
\begin{equation}
a_i^{(S,I)}=-\frac{1}{8\pi(M_1+m_2)}T^{(S,I)}_{ii} (s=(M_1+m_2)^2).
\end{equation}

First, we perform fits to the 15 LQCD data\footnote{Unless otherwise specified, to ensure that the NLO (U)ChPT is
applicable to the LQCD data, we restrict ourselves to the LQCD data obtained with $m_\pi$ ranging from 301 to 510 MeV and excluding
the heaviest point of $m_\pi=611$ MeV.}
 with the NLO HM ChPT and covariant ChPT. The $\mathring{M}$ appearing in the HQS loop function of Eq.~(\ref{GHQS}) is set equal to $\mathring{m}_D$ for the $D(D^*)$ sector and $\mathring{m}_B$ in the $B(B^*)$ sector. The results are shown in Table \ref{tab:fit1}. It seems that
both approaches fail to achieve a $\chi^2/\mathrm{d.o.f.}$ of about 1, but the
covariant ChPT describes the LQCD data better than the HM ChPT. The smaller $\chi^2/\mathrm{d.o.f.}$ in the covariant ChPT should be
attributed to the terms with the coefficients, $c_4$ and $c_5$. These two terms  cannot be distinguished from the terms with coefficients $c_2$ and $c_3$ in the HM ChPT, as mentioned earlier.

Next we perform fits using the NLO HM  UChPT and the covariant UChPT, with the loop function in the latter regularized  in either  the HQS scheme or the $\chi$-SU(3) scheme.
The results are shown in Table \ref{tab:fit2}.
 A few points are noteworthy. First, the NLO UChPT describes the LQCD data better than the NLO ChPT.
Second,  the covariant UChPT describes the LQCD data much better than the HM UChPT. The $\chi$-SU(3) approach gives a $\chi^2/\mathrm{d.o.f.}$ in between those of the HM UChPT and
the covariant UChPT.

These results are consistent with the findings from the studies of
 the decay constants of the heavy-light mesons~\cite{Geng:2010df} and the ground-state octet baryon masses
in the one-baryon sector~\cite{MartinCamalich:2010fp}. That is to say, the covariant ChPT appears to be superior in describing the light-quark mass evolution of
physical observables as compared to its nonrelativistic counterpart.

\begin{figure}
\includegraphics[width=1.0\textwidth]{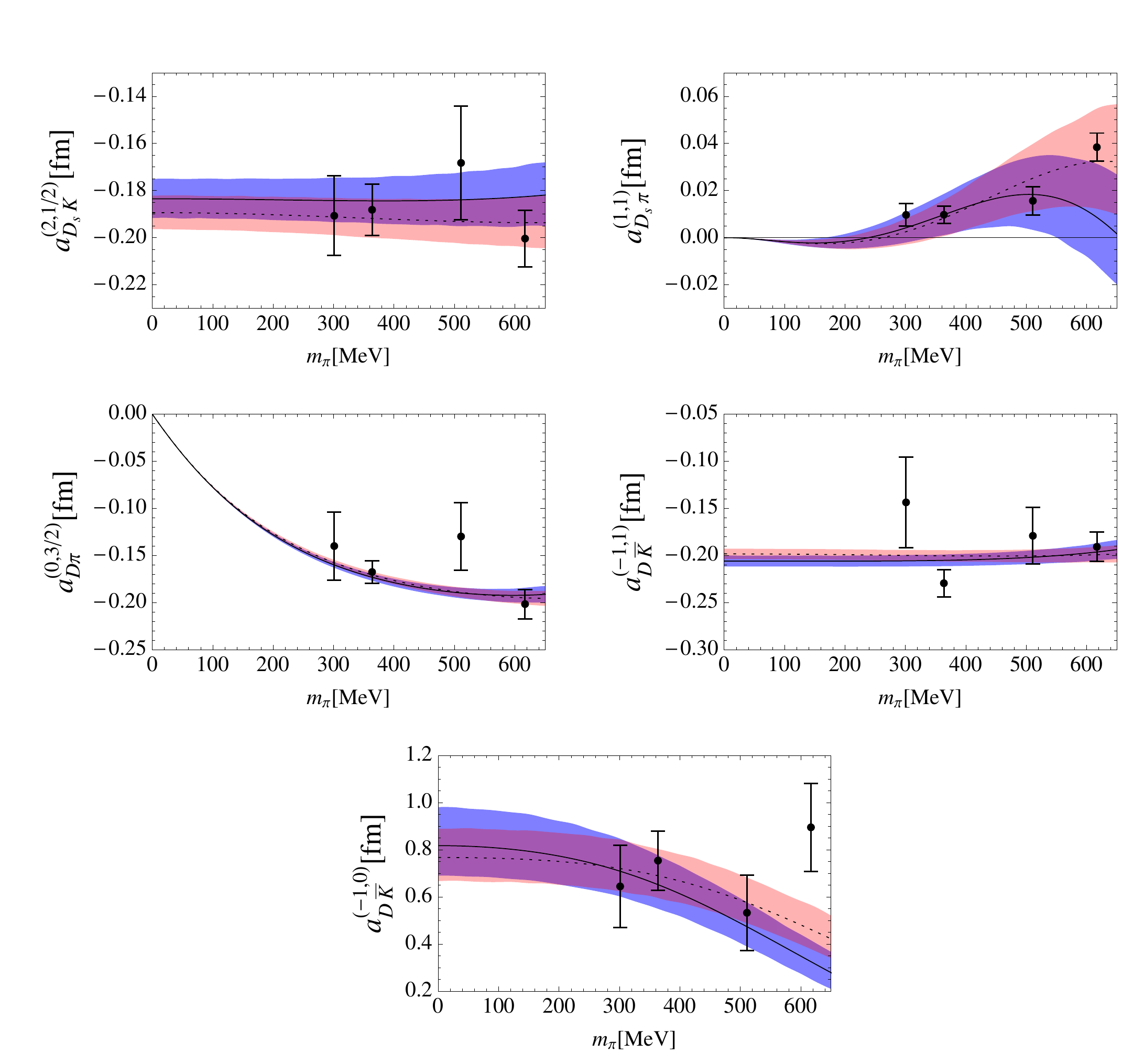}
\caption{The $n_f=2+1$ LQCD data~\cite{Liu:2012zya} vs the NLO covariant UChPT. The black solid and dashed lines show the best fits to the 15 LQCD points and  to the 20 LQCD points, with the blue and red bands covering the uncertainties propagated from
those of the LECs within one standard deviation, respectively. \label{fig:lat}}
\end{figure}

In Fig.~\ref{fig:lat},  the LQCD data are contrasted with the NLO covariant UChPT. The theoretical bands are generated from the uncertainties of the LECs.
The $D$ ($D_s$) masses are described with the NLO mass relations of Eqs.~(\ref{MassForm1}) and (\ref{MassForm2}), where the LECs $m_D$, $c_0$, and $c_1$ are fixed by fitting to the LQCD masses of
Ref.~\cite{Liu:2012zya}. In addition, the kaon mass is expressed as $m_K^2=a m_\pi^2+b$ with $a$ and $b$  determined by the LQCD data of Ref.~\cite{Liu:2012zya} as well. However, one should notice
that such a comparison is only illustrative because the NLO mass formulas cannot describe simultaneously both the LQCD $D$ and $D_s$ masses and their experimental counterparts, as also noticed
in Ref.~\cite{Liu:2012zya}.
In fact, the $\chi^2/\mathrm{d.o.f.}$ shown in Tables \ref{tab:fit1} and \ref{tab:fit2} are calculated with the $D$ and $D_s$ mass data taken directly from LQCD and not with the fitted masses of the NLO ChPT. For the sake of comparison, we show also in Fig.~\ref{fig:lat} the theoretical results obtained from a fit to all of the 20 LQCD data.  Within uncertainties they tend to overlap
with those calculated with the LECs from the fit to the 15 LQCD points.
\begin{table*}[t]
      \renewcommand{\arraystretch}{1.2}
     \setlength{\tabcolsep}{0.1cm}
     \centering
     \caption{\label{tab:fit1} Low-energy constants and the $\chi^2/\mathrm{d.o.f.}$ from the best fits to the LQCD data~\cite{Liu:2012zya} in  the covariant ChPT and the HM ChPT up to NLO, where $c_{24}=c_2-2c_4$ and $c_{35}=c_3-2c_5$. The uncertainties
     of the LECs given in the parentheses correspond to one standard deviation.}
     \begin{tabular}{lccccc}
     \hline\hline
     & $c_{24}$ & $c_{35}$ & $c_4$ & $c_5$ & $\chi^2/\mathrm{d.o.f}$ \\
    Covariant ChPT  & 0.153(35) & $-0.126(71)$ & 0.760(186) & $-1.84(39)$ &  2.01\\
    HM ChPT  & 0.012(6) & $0.167(17)$ & $\cdots$ & $\cdots$ &  3.10\\\hline\hline
    \end{tabular} % \par
\end{table*}
\begin{table*}[t]
      \renewcommand{\arraystretch}{1.2}
     \setlength{\tabcolsep}{0.1cm}
     \centering
     \caption{\label{tab:fit2} Low-energy constants, the subtraction constants, and the $\chi^2/\mathrm{d.o.f.}$ from the best fits to the LQCD data~\cite{Liu:2012zya} in the
     HQS UChPT, the $\chi$-SU(3) UChPT, and the HM UChPT. The renormalization scale $\mu$ is set at 1 GeV. The uncertainties of the LECs given in the parentheses  correspond to one standard deviation.}
     \begin{tabular}{ccccccc}
     \hline\hline
     & $a$ & $c_{24}$ & $c_{35}$ & $c_4$ & $c_5$ & $\chi^2/\mathrm{d.o.f}$ \\
  HQS UChPT & $-4.13(40)$ & $-0.068(21)$ & $-0.011(31)$
  & $0.052(83)$ & $-0.96(30)$ & 1.23 \\
    $\chi$-SU(3) UChPT&$\cdots$& $-0.096(19)$ & $-0.0037(340)$ & $0.22(8)$ & $-0.53 (21)$ & 1.57\\
  HM UChPT  & 2.52 (11) & 4.86(30) & $-9.45(60)$ & $\cdots$ & $\cdots$ &2.69 \\
\hline\hline
    \end{tabular} % \par
\end{table*}

\subsection{Dynamically generated heavy-light mesons}

Once the subtraction constant and the LECs are fixed, one can utilize the UChPT to study
whether the interactions between HL mesons and NGBs are strong enough to generate bound states or resonances, by searching for poles in the complex $\sqrt{s}$ plane.
We notice that the subtraction constant in the HM UChPT given in Table \ref{tab:fit2} is positive, and as a result, there is no bound
state generated in the $(S,I)=(1,0)$ channel. On the other hand, using the covariant UChPT, a bound state is found at $\sqrt{s}=2317\pm10$ MeV in the complex plane. We identify this bound state as
the $D_{s0}^*(2317)$.   In addition, one more state
is generated in the $(S,I)=(0,1/2)$ channel. All of them are tabulated in Table \ref{DPeaks}. In calculating the positions of these states, we have used the physical masses listed in Table \ref{table:par}.  The uncertainties in the positions of these states are estimated by changing the LECs and the subtraction constant within their 1$\sigma$ uncertainties given in Table \ref{tab:fit2}. Furthermore, we predict the heavy-quark spin partners of the $0^+$ states as well. The counterpart of the $D^*_{s0}(2317)$ appears at $\sqrt{s}=2457\pm17$ MeV,\footnote{The uncertainties are propagated from the uncertainties of the LECs and the subtraction constant. In addition, we have assigned a $10\%$ uncertainty for relating the LECs in the $D^*$ sector with those in the $D$ sector by use of heavy-quark spin symmetry. To relate the LECs between $D$ and $B$ sectors,  a $20\%$ uncertainty is assumed, and $m_\mathrm{sub}$ is varied from $m_\pi$ to $m_\eta$.}  which we identify as the $D_{s1}(2460)$. It is clear that the heavy-quark spin symmetry is approximately conserved in the HQS UChPT.

One appealing feature of the renormalization scheme we proposed in this work is that the heavy-quark flavor symmetry is conserved up to $1/M_\mathrm{HL}$, in contrast to the
naive $\overline{\mathrm{MS}}$ subtraction scheme. As such, we can calculate the bottom partners of the $D^*_{s0}(2317)$ and $D_{s1}(2460)$  with reasonable confidence. We tabulate in Table \ref{BPeaks} the bottom counterparts of the charm states of Table \ref{DPeaks}.  It should be noted that the absolute positions of these resonances are subject to corrections of a few tens of MeV because of the uncertainty related to the evolution of the UChPT from the charm sector to the bottom sector. On the other hand, the mass differences between the $1^+$ states and their $0^+$ counterparts should be more stable, as has been argued in a number of different studies (see, e.g., Ref.~\cite{Cleven:2010aw}).

\begin{table}[H]
      \renewcommand{\arraystretch}{1.2}
     \setlength{\tabcolsep}{0.1cm}
\caption{\label{DPeaks}Pole positions $\sqrt{s}=M-i\frac{\Gamma}{2}$ (in units of MeV) of charm mesons dynamically generated in the HQS UChPT. }
\centering
\begin{tabular}{c|c|c}
\hline\hline
  $(S,I)$ & $J^P=0^+$& $J^P=1^+$ \\
\hline
  (1,0) & $2317\pm10$ & $2457\pm17 $  \\
  (0,1/2)& $(2105\pm4)-i(103\pm7)$ & $(2248\pm6)-i (106\pm13)$\\
   \hline\hline
\end{tabular}
\end{table}

\begin{table}[htpb]
      \renewcommand{\arraystretch}{1.2}
     \setlength{\tabcolsep}{0.1cm}
\caption{\label{BPeaks}Pole positions $\sqrt{s}=M-i\frac{\Gamma}{2}$ (in units of MeV) of bottom mesons dynamically generated in the HQS UChPT.}
\centering
\begin{tabular}{c|c|c}
\hline\hline
  $(S,I)$ & $J^P=0^+$ & $J^P=1^+$  \\
\hline
  (1,0) & $5726\pm 28$ & $5778\pm26$  \\
  (0,1/2) & $(5537\pm14)-i(118\pm22)$ & $(5586\pm 16)-i(124\pm25)$ \\
   \hline\hline
\end{tabular}
\end{table}

\begin{table}[htbp]
      \renewcommand{\arraystretch}{1.2}
     \setlength{\tabcolsep}{0.1cm}
\centering
\caption{\label{tab:boom} Dynamically generated  $0^+$ and $1^+$ bottom states in $(S,I)=(1,0)$ from different formulations of the UChPT. Masses of the states are in units of MeV. }
\begin{tabular}{c|c|c|c|c}
\hline\hline
$J^P$   & Present work &  NLO HMChPT ~\cite{Cleven:2010aw} &  LO UChPT~\cite{Guo:2006fu}  & LO  $\chi$-SU(3)~\cite{Kolomeitsev:2003ac} \\ \hline
 $0^+$   & $5726\pm28$  & $5696\pm36$ & $5725\pm39$   & 5643 \\
 $1^+ $ & $ 5778\pm26$ &  $5742\pm 36$ & $5778\pm 7$ &  5690\\
  \hline\hline
\end{tabular}
\end{table}

In Table \ref{tab:boom} we compare the predicted $0^+$ and $1^+$ states from several different formulations of UChPT in the bottom sector. It is seen that the absolute positions can
differ by as much as  80 MeV, which is not surprising because the heavy-quark flavor symmetry was implemented differently.

\begin{figure}
\includegraphics[width=\textwidth]{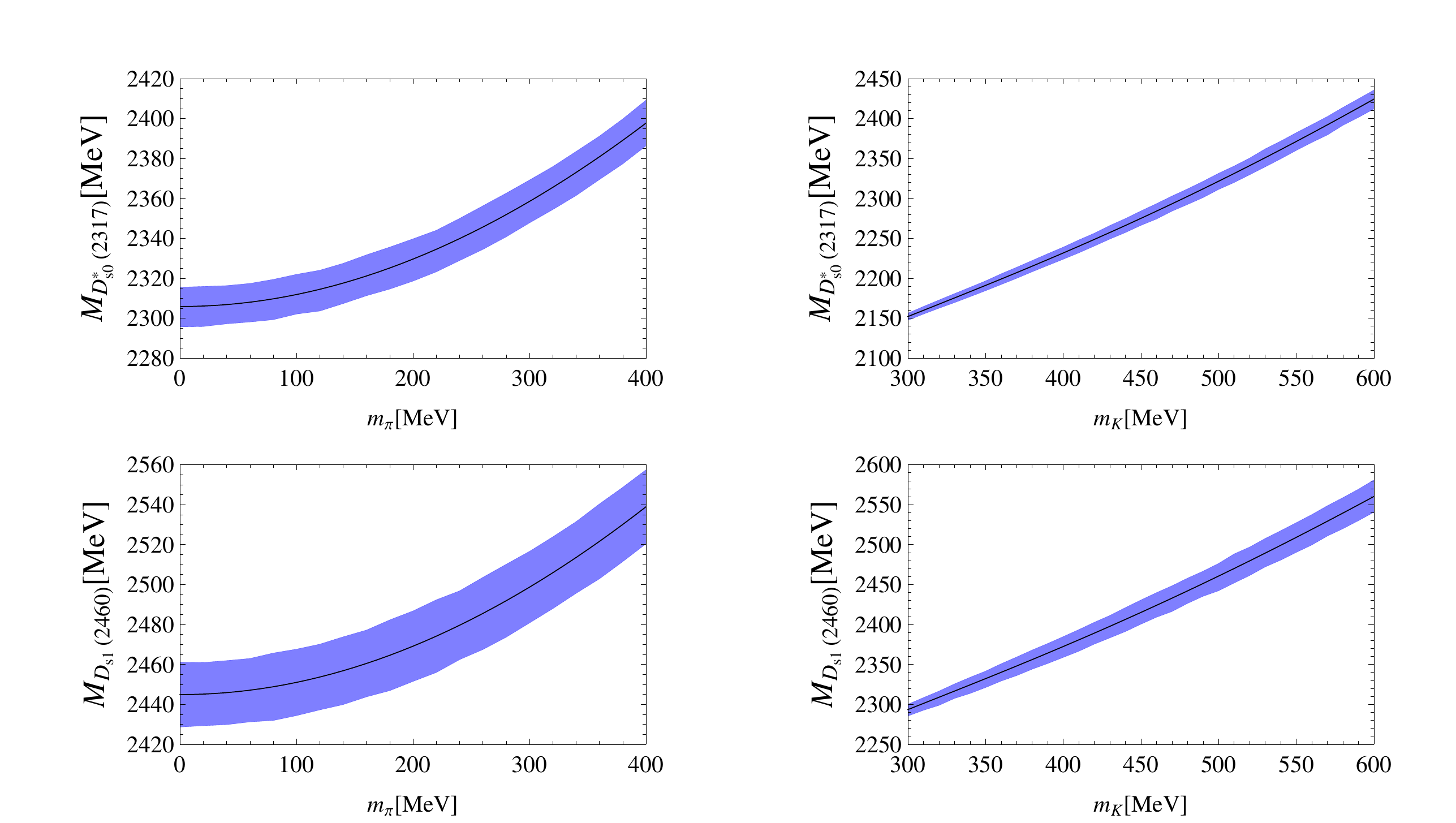}
\caption{Pion and kaon mass evolution of the pole positions of the $D^*_{s0}(2317)$ and the $D_{s1}(2460)$. \label{fig:pole}}
\end{figure}

It has been argued that the light-quark mass evolution of the masses of mesons and baryons can provide important hints about their nature (see, e.g., Refs.~\cite{Cleven:2010aw,Hanhart:2008mx}). In the left panel of Fig.~\ref{fig:pole}, we show how the pole positions of
the $D^*_{s0}(2317)$ and the $D_{s1}(2460)$ evolve as
a function of $m_\pi$. The strange-quark mass is fixed to its physical value using the leading-order ChPT. The light-quark mass dependences of the $D(D_s)$ and $D^*(D_s^*)$ are given by the NLO ChPT formulas of Eqs.~(\ref{MassForm1})-(\ref{MassForm4}). The right panel of Fig.~\ref{fig:pole} shows the evolution of the $D^*_{s0}(2317)$ and $D_{s1}(2460)$ pole position as a function of the kaon mass (or equivalently the strange-quark mass) as we fix the pion mass to its physical value. As has been
argued in Ref.~\cite{Cleven:2010aw}, the feature of being dynamically generated dictates that the dependences of the masses of these states on $m_K$ are linear with a slope close to unity, which can be clearly seen from Fig.~\ref{fig:pole}.

\section{Summary and Conclusions}
We have studied the latest fully dynamical LQCD simulations for
the scattering lengths of  Nambu-Goldstone bosons off $D$ mesons  in covariant chiral perturbation theory and its unitary version up to next-to-leading order. It
is shown that the covariant (U)ChPT describes the LQCD data better than its nonrelativistic (heavy-meson) counterpart. In addition, we show that the $D^*_{s0}(2317)$ can be dynamically generated without
\textit{a priori} assumption of its existence.

We have proposed a new subtraction scheme
to ensure that the loop function appearing in the Bethe-Salpeter equation satisfies the chiral power counting rule and has a well-defined behavior in the limit of infinite heavy-quark mass.
It is shown that this scheme has a similar $1/M_\mathrm{HL}$ scaling as the HMChPT loop function but provides a better description of the light-quark mass dependence of the LQCD scattering lengths,
in agreement with the findings in the one-baryon sector. With such a scheme, we have predicted the counterparts of the $D^*_{s0}(2317)$ in the $J^P=1^+$ sector and in the bottom sector. The experimental confirmation
of the dynamically generated states in the bottom sector can serve as a stringent test of our theoretical model and the interpretation of the $D^*_{s0}(2317)$ as a dynamically generated state from the strong $DK$ interaction.

\section{Acknowledgements}
This work is supported in part by BMBF, by the A.v. Humboldt foundation, the Fundamental Research Funds for the Central Universities, the National Natural Science Foundation of China (Grant No. 11005007),  the
New Century Excellent Talents in University Program of Ministry of Education of China under
Grant No. NCET-10-0029, the DFG  Cluster of Excellence ``Origin and Structure of the Universe,"
and by DFG and NSFC through the Sino-German CRC 110 ``Symmetries and Emergence of Structure in QCD."

\end{document}